\documentclass[a4paper]{jpconf}
\usepackage{graphicx}
\usepackage{amsmath,amsbsy,amsmath}
\begin{document}
\title{Inflationary gravitational waves\\ and exotic pre Big Bang Nucleosynthesis cosmology}

\author{A Di Marco$^{1,}$$^2$, G De Gasperis$^{1,}$$^2$, G Pradisi$^{1,}$$^2$, P Cabella}

\address{$^1$ Dipartimento di Fisica, Universit\`a degli Studi di Roma 
Tor Vergata, Via della Ricerca Scientifica, 1 I-00133 ROMA RM}
\address{$^2$ INFN Sez. di Roma 2, Via della Ricerca Scientifica, 1 
I-00133 ROMA RM}

\ead{alessandro.di.marco@roma2.infn.it}

\begin{abstract} 
According to the most popular scenario, the early Universe
should have experienced an accelerated expansion phase, called Cosmological Inflation,
after which the standard Big Bang Cosmology would have taken place giving rise to the radiation-dominated epoch.
However, the details of the inflationary scenario are far to be completely understood.  
Thus, in this paper we study if possible additional (exotic) cosmological
phases could delay the beginning of the standard Big Bang history and alter some theoretical predictions related to the inflationary cosmological perturbations,
like, for instance, the order of magnitude of the tensor-to-scalar ratio $r$.

\end{abstract}

\section{Introduction}

The evolution of the very early Universe is matter of an in-depth and long-standing debate and, to date, it is mainly the subject of speculations.  According to the standard lore, the Universe initially experienced a highly symmetric state (at the so-called Planck scale $M_p\sim 10^{18}$ GeV) 
in which the four fundamental gravitational, strong, weak and electromagnetic interactions were unified in a single fundamental force described by a fundamental (and still unknown) quantum gravity theory. Due to the expansion and the consequent cooling of the Universe, 
at some lower energy scales the gravitational interaction decoupled and the Universe entered an hypothetical phase where only strong, weak and electromagnetic interactions were unified\footnote{Unification is not necessary and does not play a fundamental role in what follows.} 
in a single force, characterized by a single coupling constant, and, eventually, with the presence of supersymmetry.
This phase is commonly known as Grand Unified epoch and it is described by a corresponding Grand Unified (gauge) Theory (GUT)
based, for instance, on a gauge group like $SU(5)$ or $SO(10)$, 
that contains the gauge group of the Standard Model of particle physics \cite{1,2,3}. 
At an energy scale of the order, presumably, of $M_{GUT}\sim 10^{16}$ GeV, a spontaneous symmetry breaking (SSB) of the GUT itself took place and the interactions separated into those of the Standard Model, described by the direct-product group $SU(3)^{c}\times SU(2)\times U(1)$ related to the strong, weak and electromagnetic interactions, respectively. 
As the temperature of the Universe decreased, the three different coupling constants ran back in different ways. 
Finally, at the energy scale of around $\sim 10^{2} \div 10^3$ GeV the electroweak symmetry breaking\footnote{If supersymmetry is present, one has also to 
accomodate its breaking at a scale higher than the LHC limits.}
took also place and $SU(3)^{c} \times U(1)^{em}$ survived as the residual exact gauge symmetry of the present Universe, described by the well-known Standard Model vacuum phase. 

However, cosmological and astrophysical observations suggest that the early Universe also underwent an almost de Sitter-like phase, called Inflation \cite{4,5,6}.
It occurred likely at an energy scale intermediate between the GUT phase transition and the electroweak scale (say at $M_{inf} \leq 10^{16}$ GeV) 
and made the Universe extremely large, flat, isotropic and homogeneous on astronomical scales, 
as well as equipped with cosmological scale-invariant metric perturbations of two types.
The first type are scalar perturbations, that seeded the formation of large scale structures (LSS)
and led to the cosmic microwave background (CMB) temperature anisotropies.  
The second are tensor perturbations, i.e. primordial gravitational waves (GW), 
that represent a distinctive signature of inflation and whose detection\footnote{The detection of inflationary GW can be direct, as happens for the gravity waves from astrophysical merging events,
or indirect, thanks to the polarization induced on the CMB.} 
would confirm the goodness of the inflationary paradigm itself.

The simplest version of inflation is the well known slow roll scenario.
In this case the mechanism 
is driven by some fundamental scalar field $\varphi$, called inflaton, minimally coupled to
gravity and slowly probing an almost flat region (that plays the role of a false vacuum) of the corresponding effective scalar potential $V(\varphi)$.

At the end of inflation, where the potential steepens, the inflaton field
falls in the true vacuum of the potential, oscillates and decays in an hot plasma constituted by the standard model (SM) relativistic particles, 
providing the reheat of the Universe and the beginning of the Big Bang evolution.
The reheating phase \cite{7,8,9} is expected to be a very complicated epoch, whose details are expected to be strongly model dependent.
Therefore, it comes not as a surprise that the energy scale at the end of the process, parametrized by the so called reheating temperature, could well be placed 
at around $10^{14}$ GeV but also at much lower scales.

It is important to stress that the mentioned chronology
basically represents just a paradigmatic scenario among the many proposed in a vast literature during the last decades.
For example, our Universe might not have experienced a GUT phase but rather an inflationary epoch triggered already at the Planck scale and accompanied by a subsequent SM phase. 
Indeed, due to the absence of experimental data, we basically ignore the relation between Inflation and GUT's or between Inflation and the SM degrees of freedom.  
Even the (effective) quantum gravity scale could be well below the Planck scale, as it happens in some models inspired by Superstring Theory
\cite{10}.

\section{The Universe's history before the Big Bang Nucleosynthesis}

The evolution of the Universe after the hypothetical inflationary/reheating era represents a further conundrum,
because the underlying physics is not under control being the energy scale typically above the TeV scale at which the Electroweak (EW) and the Quantum Chromodynamics (QCD) phase transitions take place.

In this framework, an early post-inflationary radiation dominated Universe would be the expected standard scenario but not the unique viable possibility. Indeed, there is room for non trivial evolution in the history of the Universe immediately after the reheating era.  
In particular, the expansion of the Universe just after the reheating could have been subjected to additional phases that can be driven by one, or even more new exotic scalar components whose energy density is initially dominant over the radiation energy density.  
With the evolution, due to the dilution property of scalars and to the decreasing of the energy scale below some critical value, the radiation component becomes dominant and the most familiar Big Bang evolution takes place.

The origin of the involved (exotic) scalar fields can be identified with the fact that they are almost ubiquitous in supergravity and in superstring (orientifold) inspired models \cite{11,12,13,14}.  
One possibility is that they can be produced during reheating by the perturbative decay of the inflaton field (if it couples suitably to them) or by other, more complicated, mechanisms.  
In general, they could interact or be sterile with respect to the degrees of freedom of the Standard Model sector.  
In the last case, they can be frozen and simply dilute faster than radiation during the cosmological evolution. It is then natural to describe them in terms of a perfect fluid.  In the former case, they could decay into radiation providing a second period of reheating of the Universe, at lower energy scales.

In this document, we want to focus on some of the cosmological implications of sterile post inflationary scalar fields/fluids \cite{14,15,16}.
First of all, the introduction of this new kind of scalar degrees of freedom in the cosmological particle spectrum
requires a re-definition of the energy density of the Universe, that can be generically written as
\begin{eqnarray}
\rho(T)=\rho_r(T)+\sum_i \rho_{\phi_i}(T)=\rho_r(T)\eta(T) ,
\end{eqnarray}
where
$T$ labels the temperature scale of the Universe that we assume in the range $T_{BBN}<T<T_{reh}$.
The first term of the inequality is of fundamental importance in order not to ruin the excellent predictions of the Big Bang Nucleosynthesis (BBN) on the production of the light elements abundances.
Instead, the parameter $\eta$ describes the relative dominance of the scalar sector over the relativistic cosmological sector
\begin{eqnarray}
\eta(T)= 1 + \frac{\sum_i \rho_{\phi_i}(T)}{\rho_r(T)}.
\end{eqnarray}
Although the case of multiple scalars would be very interesting \cite{15}, 
we focus on a post reheating dynamics dominated by a fluid related to a single additional scalar field $\phi$
that dominates and then leaves room to the radiation contribution at a given transition temperature $T_*$ \cite{14,15,16}, i.e. 
\begin{eqnarray}
\rho_{\phi}>\rho_r \mbox{ for } T>T_* , \\
\rho_{\phi}=\rho_r \mbox{ for } T=T_* , \\
\rho_{\phi}<\rho_r \mbox{ for } T<T_* .
\end{eqnarray} 
The dynamics of the radiation energy density can be written in the form
\begin{equation}
\rho_r(T)= \rho_r(T_*) \frac{g_E(T)}{g_E(T_*)}\left(\frac{T}{T_*} \right)^4 , 
\end{equation}
where $g_E$ represents the number of the relativistic degrees of freedom in the context of energy defined as
\begin{equation}
g_{E}(T)=\sum_{b} g_{b}\left( \frac{T_b}{T}\right)^4 + \frac{7}{8}\sum_{f} g_{f}\left(\frac{T_f}{T}\right)^4,
\end{equation}
with $b$ and $f$ labeling bosonic and fermionic contributions, respectively.
The energy density temperature dependence of the field $\phi$ satisfies the law
\begin{equation}
\rho_{\phi}(T)=\rho_{\phi}(T_*)\left( \frac{g_S(T)}{g_S(T_*)}\right)^{\frac{4+n}{3}}\left(\frac{T}{T_*}\right)^{4+n}
\end{equation}
and $g_S$ is now the number of relativistic degrees of freedom contributing to the entropy
\begin{equation}
g_{S}(T)=\sum_{b} g_{b}\left( \frac{T_b}{T}\right)^3 + \frac{7}{8}\sum_{f} g_{f}\left(\frac{T_f}{T}\right)^3,
\end{equation}
while the index $n$ is the ``dilution" coefficient that parametrizes the deviation from the standard radiation coefficient.
By assuming that $g_S\sim g_E$ at very high energy scales (above the QCD phase transition $\sim 150-170$ GeV), one can easily obtain that \cite{14,15}
\begin{equation}\label{eqn: eta}
\eta(T)\sim 1 + \left(\frac{T}{T_*}\right)^n . 
\end{equation}
As we will see, this quantity will be of crucial importance for the subsequent analysis.

\section{Relic gravity waves and non trivial post reheating scalar dominance}

A fundamental parameter of the slow roll inflationary dynamics is the number of $e$-folds, $N$, that counts the number of exponential expansions of the Universe.
Thus, $N$ describes the duration of the de Sitter-like stage, and consequently, it determines
the theoretical predictions related to the
inflationary cosmological perturbations like for instance, the well known scalar spectral index $n_s$ and the tensor-to-scalar ratio $r$.
In general, this number depends on the post inflationary evolution of the Universe and can be approximated as (see for example \cite{14,15})
\begin{equation}\label{eqn: standard}
N\sim 64 - \frac{1-3w_{reh}}{3(1+w_{reh})}\ln\left(\frac{M_{inf}}{T_{reh}}\right) + \ln\left(\frac{M_{inf}}{M_p}\right) ,
\end{equation}
where $w_{reh}$ is the mean value of the equation of state (EoS) parameter of the ``reheating fluid", $M_{inf}$ is the energy scale at which Inflation takes place and $T_{reh}$ is the scale at which the Universe gets reheated. 
It is important to outline that it is a common lore to assume the value of the inflationary $e$-folds in the range between $50$ and $60$. 
However, the presence of an additional exotic cosmological phase after reheating and before the radiation dominated era,  
gives rise to a modification of the standard results for $N$.

\begin{figure}[htbp]
\centering
\includegraphics[width=12cm, height=8cm]{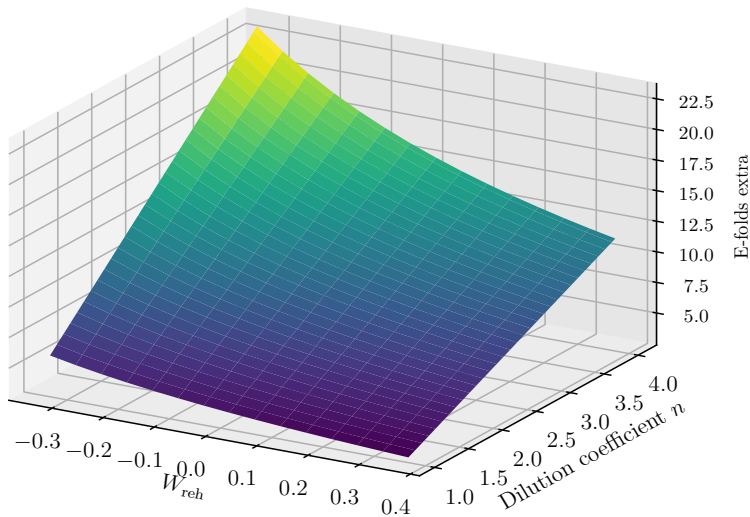}
\caption{Extra number of $e$-foldings, induced by the exotic scalar field epoch, as function of the reheating EoS, $w_{reh}$ and the constant $n$.
The reheating temperature is $\sim 10^9$ GeV while the scalar-radiation transition temperature is $\sim 10^4$ GeV.}
\label{fig: 1}
\end{figure}

This change can be characterized as $N=N_{standard}+N_{extra}$ where the extra term reads \cite{15}
\begin{equation}\label{eqn: general}
%N\quad\rightarrow\quad N=N_{standard}+N_{extra}, \quad 
N_{extra}=\frac{1}{3(1+w_{reh})}\ln\eta(T_{reh}) 
\end{equation}
where now $N_{standard}$ is just the contribution in Eq.($\ref{eqn: standard}$). Eq.($\ref{eqn: general}$) suggests how the additional epoch will provide a general modification of $N$ that depends on the magnitude of the previous parameter $\eta$ evaluated at the reheating scale.
In Fig.(1) we provide a numerical estimate of the extra $e$-folds contribution in Eq.($\ref{eqn: general}$),
as a function of the EoS parameter and of the constant $n$.
It is crucial to observe how the non standard $N_{extra}$ term affects the sector of inflationary perturbations.
To make an example, we can focus on the gravitational waves, namely on the tensor-to-scalar ratio $r$.
In many viable slow-roll inflationary models, the tensor-to-scalar ratio depends on the inverse squared of $N$.
For instance, in the simplest case of the so-called $\alpha$-attractor models \cite{17,18,19,20,21,22} in which the potential takes on the form
$V(\varphi)=M^4_{inf}(1-\exp{(-b\varphi/M_p)})^2$ with $b=\sqrt{2/(3\alpha)}$, one has
\begin{equation}\label{eqn: r}
r(N)\sim \frac{12\alpha}{N^2} , 
\end{equation}
where $\alpha$ is a constant that depends on certain specific features of the model and basically can be any positive number.
The impact of the post inflationary scalar fluid on $r$ can be thus 
derived by combining Eq.($\ref{eqn: eta}$), Eq.($\ref{eqn: standard}$) and Eq.($\ref{eqn: general}$) with Eq.($\ref{eqn: r}$). 
They tell that the tensor-to-scalar ratio assumes the following approximate form
%included through a \textbf{correction} term $r(N)=r(N_{standard})+\Delta r(N_{extra})$
\begin{equation}
r\sim 12\alpha\left[64 - \frac{1-3w_{reh}}{3(1+w_{reh})}\ln\left(\frac{M_{inf}}{T_{reh}}\right) + \ln\left(\frac{M_{inf}}{M_p}\right)
+ \frac{1}{3(1+w_{reh})}\ln\left(\frac{T_{reh}}{T_*}\right)^n \right]^{-2}.
\end{equation}
It is also very interesting to analyze the behavior of $r$ in terms of the (mean) reheating properties and of the diluting field.
This should be useful to get an idea about the order of magnitude of the enhancement ({or the reduction) of the tensor-to-scalar ratio.
To this respect, we report in Fig.(2) $r$ both as a function of $w_{reh}$ and of $n$, for the case $\alpha=1$ related to the Starobinsky inflationary model \cite{19}.
As one can appreciate, values $w_{reh}>0$ of the EoS parameter and large values of constant $n$ allow for a very little tensor-to-scalar ratio ($r<0.003$)
while little values both of the EoS and of $n$ provide values of $r$ even much larger than $0.0045$.
It is important to point out that the prediction of the Starobinsky model about $r$
is still in great agreement with the current bound ($r<0.064$, $95\%$, TT,TE,EE,lowE,lensing+BK14) provided by the Planck and BICEP2 collaboration (see \cite{23} for
details).
Actually, the enhancement induced by an additional post-reheating component provides the possibility to detect primordial inflationary GW 
on more accessible scales.

\begin{figure}[htbp]
\centering
\includegraphics[width=12cm, height=8cm]{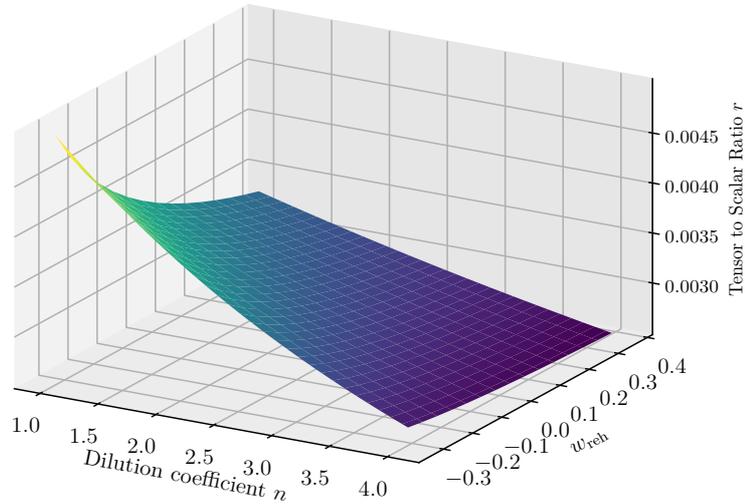}
\caption{Tensor-to-scalar ratio for a Starobinsky-like model as function of the $w_{reh}$ and $n$ with $T_{reh}\sim 10^9$ GeV and $T_{*}\sim 10^4$ GeV.}
\label{fig: 2}
\end{figure}

\section{Discussion and conclusions}

The introduction of an additional, pre-BBN, non trivial post-inflationary epoch, can provide non-negligible implications on the inflationary Universe.
In particular, it is possible to put bounds on the reheating scale of the Universe after inflation \cite{15}, as well as to
extend in a non trivial way the inflationary predictions related to the cosmological fluctuations produced during inflation, 
at least within certain classes of inflationary models \cite{14,15}.

In this paper, we have discussed how the combination of the reheating properties and of the presence of a single scalar sterile post inflationary field
can alter in a significant way the tensor-to-scalar ratio associated to the inflationary gravity waves.
As an example, we have considered the Starobinsky inflationary scenario, but of course it is possible to extend the procedure to many other classes of inflationary models.
We are convinced that the inclusion of additional scalar fields obeying a hierarchy of dilution laws ($\rho_{\phi_1}<\rho_{\phi_2}<\rho_{\phi_3}<...<\rho_{\phi_k}$) can even be quite interesting, as proposed in \cite{15}.  
The related consequences will be deepened in a future publication.  Furthermore, it is also important to observe that the presence of this kind of scalars could also affect the physics at lower energy scales, 
like for instance the freezing mechanism of dark matter particles \cite{16}.


\section*{References}
\begin{thebibliography}{10}

% Grand Unified refs
\bibitem{1} Georgi H, Glashow S L, 1974 {\it Phys. Rev. Lett.} {\bf 32} 438
\bibitem{2} Langacker, 1981 {\it Phys. Rept.} {\bf 72} 185
\bibitem{3} W. de Boer, 1994 {\it Prog. Part. Nucl. Phys.} {\bf 33} 201 (hep-ph/9402266)
%Inflation refs
\bibitem{4} Guth A 1981 {\it Phys. Rev.} D {\bf 23} 2
\bibitem{5} Linde A D 1981 {\it Phys. Lett.} B {\bf 108} 6 389-393; Albrecht A, Steinhardt P J 1982 {\it Phys. Rev. Lett.} {\bf 48}, 1220
\bibitem{6} Linde A D 1983 {\it Phys. Lett.} B {\bf 129} 3 4
%Reheating refs
\bibitem{7} Albrecht A, Steinhardt P J, Turner M S, Wilczek F 1982 {\it Phys. Rev. Lett.} {\bf 48} 20; 
Dolgov A D, Linde A D, 1982 {\it Phys. Lett.} B {\bf 116} 329;
Abbott L F, E. Farhi, Wise M B 1982  {\it Phys. Lett.} B {\bf 117} 29;
Turner M S, 1983 {\it Phys. Rev.} D {\bf 28} 1243
\bibitem{8} Dolgov A D, Kirilova D P 1990 {\it Sov. J. Nucl. Phys.} {\bf 51} 172; 
Traschen J H, Brandenberger R H 1990 {\it Phys. Rev.} D {\bf 42} 2491;
Kofman L, Linde A D, Starobinsky A A 1994 {\it Phys. Rev. Lett.} {\bf 73} 3195; 
Shtanov Y, Traschen J H, Brandenberger R H 1995 {\it Phys. Rev.} D {\bf 51} 5438;
L. Kofman, A. D. Linde and A. A. Starobinsky 1997 {\it Phys. Rev.} D {\bf 56} 3258
\bibitem{9} Bassett B A, Tsujikawa S, Wands D 2006 {\it Rev. Mod. Phys.} {\bf 78} 537; 
Allahverdi R, Brandenberger R, Cyr-Racine F Y, Mazumdar A 2010 {\it Ann. Rev. Nucl. Part. Sci.} {\bf 60} 27; 
Amin M A, Hertzberg M P, Kaiser D I, Karouby J 2014 {\it Int. J. Mod. Phys.} D {\bf 24} 1530003; 
Lozanov D K, arXiv:1907.04402
%String scale fields refs
\bibitem{10} Arkani-Hamed N, Dimopoulos S, Dvali G R, {\it Phys. Lett.} B {\bf 429} 263;
Antoniadis I, Arkani-Hamed N, Dimopoulos S, Dvali G R, {\it Phys. Lett.} B {\bf 436} 257;
Antoniadis I, Sagnotti A, {\it Class. Quant. Grav.} {\bf 17} 939
%String scalar fields refs
\bibitem{11} Angelantonj C, Sagnotti A 2002 {\it Phys. Rept.} {\bf 371} 1 Erratum: [{\it Phys.Rept.} {\bf 376} 6 407]
\bibitem{12} Blumenhagen R, Kors B, Lust D, Stieberger S 2007 {\it Phys. Rept.} {\bf 445} 1 
\bibitem{13} Dudas E 2000 {\it Class. Quant. Grav.} {\bf 17}, R41 
%Inflation refs sterile scalar fields
\bibitem{14} Maharana A, Zavala I 2018 {\it Phys. Rev.} D {\bf 97} 12 123518
\bibitem{15} Di Marco A, Pradisi G, Cabella P 2018 {\it Phys. Rev.} D {\bf 98} 12 123511
\bibitem{16} D'Eramo F, Fernandez N, Profumo S 2017 {\it J. Cosmol. Astropart. Phys.} {\bf 1705} 05 012 
\bibitem{17} Kallosh R., Linde A 2013 {\it J. Cosmol. Astropart. Phys.} {\bf 07} 002; 2013
{\it J. Cosmol. Astropart. Phys.} {\bf 12} 006
\bibitem{18} Ferrara S, Kallosh R, Linde A, Porrati M 2013 {\it Phys. Rev.} D {\bf 88} 085038 
\bibitem{19} Kallosh R, Linde A 2013 {\it J. Cosmol. Astropart. Phys.} {\bf 06} 028.
\bibitem{20} Kallosh R, Linde A, Roest D 2013 {\it J. High Energy Phys.} {\bf 11} 198.
\bibitem{21} Galante M, Kallosh R, Linde A, Roest D 2015 {\it Phys. Rev. Lett.} {\bf 114} 141302
\bibitem{22} Kallosh R, Linde A 2016 {\it J. Cosmol. Astropart. Phys.} {\bf 06} 047
\bibitem{23} Planck 2018 results. X. Constraints on inflation.
%Planck 2015 results. XIII. Cosmological parameters 2016 A&A 594 A13































\end{thebibliography}
\end{document}